\documentclass{aa}
\usepackage{txfonts}
\usepackage{graphicx}
\usepackage{subcaption}
\usepackage{threeparttable} 
\usepackage[colorlinks=true,allcolors=blue]{hyperref}
\usepackage{gensymb}

\DeclareUnicodeCharacter{5415}{}  
\DeclareUnicodeCharacter{5EFA}{}  
\DeclareUnicodeCharacter{4F1F}{}  

\begin{document} 

\authorrunning{Mountrichas et al.}
\titlerunning{Halo mass, star formation, and morphology}

\title{Halo mass, star formation, and morphology in galaxies: A morphology-dependent connection between halo environment and star formation activity}

\author{G. Mountrichas\inst{1}, F. Shankar\inst{2}, F. J. Carrera\inst{1},  A. Georgakakis\inst{3} }
          
     \institute {Instituto de Fisica de Cantabria (CSIC-Universidad de Cantabria), Avenida de los Castros, 39005 Santander, Spain.
              \email{gmountrichas@gmail.com}
               \and
             Department of Physics and Astronomy, University of Southampton, Highfield, Southampton, SO17 1BJ, UK
              \and
              Institute for Astronomy \& Astrophysics, National Observatory of Athens, V. Paulou \& I. Metaxa, 11532, Greece
             }

\abstract
{Understanding the mechanisms responsible for the quenching of star formation in galaxies requires disentangling the roles of internal galaxy structure and the large-scale environment.}
{We investigate how galaxy morphology and the specific star formation rate (sSFR) relate to the typical masses of host dark-matter halos (DMHs), with the aim of determining whether the connection between halo environment and star formation activity depends on galaxy morphology.}
{We analysed non-active galaxies in the VIPERS and Stripe 82 fields at $0.5 \leq z \leq 1.2$, combining DESI-based S\'ersic classifications with stellar masses ($M_\star$) and star formation rates (SFRs) derived from spectral energy distribution fitting. We separated galaxies into disc- and bulge-dominated systems and divided them into quantile-based sSFR bins designed to minimise $M_\star$ differences between morphologies. To account for the redshift evolution of the star-forming main sequence, we characterised each bin by its median $\Delta{\rm MS} = \log {\rm SFR} - \log {\rm SFR}_{\rm MS}$. We measured projected galaxy cross-correlation functions and inferred characteristic DMH masses from the clustering amplitudes.}
{The relation between halo mass and star formation activity depends strongly on morphology. Disc-dominated galaxies exhibit little variation in halo mass across the range of $\Delta{\rm MS}$ probed, whereas bulge-dominated galaxies occupy lower-mass halos at the high-$\Delta{\rm MS}$ end and higher, mutually consistent halo masses towards lower $\Delta{\rm MS}$. These trends are unlikely to be driven primarily by $M_\star$ differences, which remain modest within and across the bins. At the same time, all subsamples already inhabit relatively massive halos, while many galaxies remain on or above the star-forming main sequence.}
{Typical halo mass alone does not map cleanly onto quiescence in the present sample. Instead, our results suggest a morphology-dependent connection between halo environment and star formation activity, with a stronger link between the two in bulge-dominated galaxies than in disc-dominated systems.}

   
\maketitle  

\section{Introduction}

Understanding how galaxy properties relate to their large-scale
environment is central to studies of galaxy formation and evolution.
Clustering measurements provide a direct way to connect observable
galaxy properties to the masses of their host dark-matter halos (DMHs),
thereby linking internal galaxy processes to the growth of cosmic
structure.
It is now well established that galaxy clustering relates strongly to the 
stellar mass ($M_\star$) and star formation activity, with more massive
and quenched systems residing, on average, in more massive halos
(e.g. \citealt{Zehavi2011, Coupon2015, Mountrichas2019}).
From a physical perspective, these trends reflect the combined
influence of internal processes linked to galaxy mass and structure,
such as feedback and morphological stabilisation, and external
halo-scale mechanisms that regulate the supply of cold gas through
shock heating and maintenance heating of the circumgalactic medium
(e.g. \citealt{Peng2010, Keres2005, DekelBirnboim2006, Croton2006,
GaborDave2012}).

Beyond $M_\star$ and the star formation rate (SFR), galaxy morphology
encodes key information about a galaxy’s assembly history and
dynamical state.
Bulge-dominated systems are often associated with quenched star
formation and dense environments, while disc-dominated galaxies
preferentially inhabit lower-density regions.
These trends are encapsulated in the classic morphology--density and
morphology--clustering relations observed in the local and
intermediate-redshift Universe
(e.g. \citealt{Dressler1980, PostmanGeller1984, Skibba2009,
Wilman2010, Coil2017}).
However, disentangling the relative roles of morphology, star
formation, and environment remains challenging, as these properties
are strongly correlated.

A large body of previous work has explored the connection between
star formation activity and environment, showing that quenched or red
galaxies preferentially reside in more massive DMHs
(e.g. \citealt{Peng2010, Tinker2013, Moustakas2013,
ZuMandelbaum2016}).
In parallel, theoretical and semi-analytic studies have proposed
morphological quenching scenarios, in which the presence of a
dynamically hot stellar component or a dominant bulge can suppress
star formation without requiring a substantial change in halo mass
(e.g. \citealt{Martig2009, Martig2013}).
While these studies have provided important insights, observational
efforts have typically focused either on star formation activity or
morphology in isolation, or have relied on local density measures
rather than direct estimates of halo mass, making it difficult to
isolate the relative impact of halo mass and internal galaxy structure
on the quenching of star formation. Very recent work has further strengthened the case for halo
quenching, arguing that halo mass may be the dominant prerequisite
for quenching in central galaxies once biases in halo-mass estimates
are carefully accounted for \citep{ZhaoPeng2025}.
At the same time, comparisons between observations and simulations
continue to highlight tensions in reproducing the joint dependence
of quiescence, halo mass, environment, and active galactic nucleus (AGN) activity
\citep{YesufBottrell2026}.

At the same time, a number of studies have argued that quenching
correlates at least as strongly with central galaxy properties such as
bulge mass, central stellar density, and black-hole mass as with halo
mass itself (e.g. \citealt{Woo2013, Bluck2014, Bluck2016,
Piotrowska2022, Goubert2025}). It therefore remains an open question
whether halo mass is the primary driver of quenching or instead traces
other aspects of central galaxy growth that are more directly linked to
the shutdown of star formation. A further complication is that the relative roles of halo environment
and internal structure may differ between central and satellite
galaxies, whose quenching pathways need not be the same
(e.g. \citealt{Peng2012, Woo2015, Knobel2015}).

In a series of recent papers, we investigated the connection between
galaxies, AGNs, and their host DMHs using
large-area X-ray surveys and galaxy samples.
In Paper~I \citep{Mountrichas2026a}, we explored AGN clustering as a function of black-hole
mass, Eddington ratio, and X-ray luminosity.
In Paper~II \citep{Mountrichas2026b}, we linked black-hole scaling relations to large-scale
structure, demonstrating that environmental differences emerge
primarily at the high-mass end of the supermassive black hole (SMBH) population.
In both studies, host-galaxy properties such as $M_\star$ and SFR were
shown to play a dominant role in shaping clustering trends.

This paper is the third in our clustering-based investigation of halo
mass, galaxy properties, and black hole activity, while remaining
self-contained through its specific focus on the non-AGN galaxy
population.
In this paper (Paper~III), we investigate how the large-scale
environments and typical host halo masses of non-AGN galaxies vary with
morphology and the specific star formation rate (${\rm sSFR} =
{\rm SFR}/M_\star$).
Using DESI-based morphological measurements and homogeneous $M_\star$
and SFR estimates, we classify galaxies by Sérsic index and split them
into sSFR bins defined via a quantile-based scheme.
We then measure galaxy clustering in each subsample and infer the
typical DMH masses as a function of both morphology and
star formation activity.

This approach allows us to address whether, at a fixed sSFR,
galaxies of different morphologies inhabit different DMHs.
Conversely, we ask whether the well-known environmental dependence on
star formation activity is equally strong for disc- and
bulge-dominated systems.
By explicitly controlling for star formation activity while
separating galaxies by morphology, and by using direct halo mass
estimates from clustering, our analysis provides morphology-resolved empirical constraints on
how halo environment and star formation activity are connected across
different structural galaxy populations.
Throughout this paper, we adopt a flat $\Lambda$ cold dark matter ($\Lambda$CDM) cosmology with 
$\Omega_{\mathrm{M}}=0.315$, $\Omega_{\Lambda}=0.685$, $h=0.674$ (i.e.\ $H_0 = 67.4~\mathrm{km\,s^{-1}\,Mpc^{-1}}$), and 
$\sigma_8(z=0) = 0.811$
consistent with the \citet{Planck2020} cosmological parameters.

\section{Data and host-galaxy properties}
\label{sec_data}

In this section we present the galaxy datasets used in our analysis. We also describe the SED fitting analysis used to calculate the galaxy properties.

\subsection{Galaxy samples}

The galaxy samples used in our analysis were drawn from the second public data release of the VIPERS (PDR-2; \citealt{Scodeggio2018}) and Stripe~82 \citep[][]{Jiang2014} fields, which were both also used in Papers~I and II. In VIPERS, the clustering analysis was based on galaxies with reliable
spectroscopic redshifts.
In Stripe~82, the galaxy sample was based primarily on spectroscopic
redshifts, supplemented where necessary by high-quality photometric
redshifts, following the same construction adopted in Papers~I and II. In both cases, we restricted the analysis to the redshift range $0.5 \leq z \leq 1.2$, ensuring consistency with our previous clustering studies.

The final galaxy samples used for clustering consist of 2428 galaxies in the VIPERS field and 19\,141 galaxies in Stripe~82.
These samples are identical to the galaxy samples employed in Papers~I and II, and we refer the reader to those works for a detailed description of survey characteristics, target selection, and completeness.

\subsection{Stellar masses and star formation rates}

Stellar masses ($M_\star$) and SFRs for all galaxies were derived through spectral energy distribution (SED) fitting using the same CIGALE-based methodology adopted in Papers~I and II \citep{Boquien2019, Yang2022}.
This homogeneous approach ensured that galaxy properties were directly comparable across fields and subsamples.

Briefly, the SED fitting used broad-band photometry from the optical to
the mid-infrared and incorporated a delayed star formation history,
supplemented by a recent burst component.
Stellar emission was modelled using the \citet{Bruzual_Charlot2003}
templates and attenuated following \citet{Charlot_Fall_2000}, while dust
emission was described using the \citet{Dale2014} models \citep[e.g.][]{Mountrichas2021b, Mountrichas2021c, Mountrichas2022c, Mountrichas2023c, Mountrichas2023d, Mountrichas2024a, Mountrichas2024b, Mountrichas2024c, Mountrichas2024d, Mountrichas2025a}.
We applied the same quality cuts as in our previous work, requiring reliable photometric coverage, acceptable reduced $\chi^2$, and consistency between best-fit and Bayesian estimates of $M_\star$ and SFR \citep[e.g.][]{Mountrichas2022a, Mountrichas2022b, Mountrichas2023a, Mountrichas2023b}.

Galaxies with a significant AGN component in their SEDs were excluded.
This choice is deliberate: the aim of the present paper is to study the
relation between halo environment, morphology, and star formation
activity in the non-active galaxy population, complementing Papers~I and
II of this series, which focus explicitly on AGNs.

We note, however, that this selection is not expected to be morphology neutral.
Because bulge-dominated galaxies host, on average, more massive black
holes, they are more likely to be identified as AGNs through SED-based
diagnostics, whereas AGNs hosted by strongly star-forming disc galaxies
are more easily diluted by host-galaxy emission.
As a result, the AGN exclusion preferentially removes a fraction of
high-concentration systems from the parent galaxy sample.
This effect is important for interpreting the comparison with previous
galaxy-only studies and is discussed further in Section~4.2.

\section{Morphological classification and sSFR binning}
\label{sec_analysis}

In this section, we describe the morphological classification of
galaxies. We also outline the binning scheme adopted to separate the
sample into bins of sSFR.

\subsection{Morphological classification}
\label{sec_morphology}

Galaxy morphology is characterised using structural parameters
from the DESI Legacy Imaging Surveys \citep{Dey2019_DESILegacy}.
We cross-matched our galaxy samples with the Legacy Surveys sweep
catalogues and adopted structural quantities derived from the Tractor
parametric surface-brightness fits.

The Sérsic-based morphology measurements and the SED-derived stellar
population properties do not rely on the same photometric inputs.
The morphological classifications were taken from DESI Legacy Survey
imaging, whereas $M_\star$ and SFR were derived independently from
broad-band optical-to-mid-infrared photometry through SED fitting.
This avoids circularity between the structural classification and the
stellar-population modelling, although differences in bandpass,
resolution, and depth between the imaging products remain an important
source of uncertainty and are discussed further in Section~5.4.

The structural parameters used in this work include the Sérsic index
(\texttt{SERSIC}), the model half-light radius (\texttt{SHAPE\_R}),
and the adopted morphological model (\texttt{TYPE}).
The \texttt{TYPE} parameter identifies the preferred Tractor model,
including point-source (\texttt{PSF}), round exponential
(\texttt{REX}), exponential (\texttt{EXP}), de Vaucouleurs
(\texttt{DEV}), and Sérsic (\texttt{SER}) profiles.
The \texttt{SHAPE\_R} parameter gives the half-light radius of the
adopted galaxy model. The \texttt{SERSIC} parameter provides the
corresponding Sérsic index, with $n=1$ for exponential
(\texttt{EXP} and \texttt{REX}) profiles and $n=4$ for
de Vaucouleurs (\texttt{DEV}) profiles, while for sources fitted
with a Sérsic model (\texttt{SER}) it gives the fitted Sérsic index.

In this work, we adopted the Sérsic index as our primary
morphological indicator.
This choice was motivated by its direct physical interpretation
as a tracer of light concentration and bulge prominence, as well
as its continuous nature, which allows for a flexible
classification of galaxy structure.
While the \texttt{TYPE} parameter provides a useful coarse
classification, it encodes morphology in discrete categories
and is more sensitive to model-selection boundaries.
Similarly, \texttt{SHAPE\_R} primarily traces galaxy size rather
than internal structure.
For these reasons, the Sérsic index provides the most suitable
and physically motivated parameter for separating disc- and
bulge-dominated systems in the context of this analysis.

We emphasise, however, that these Sérsic measurements are derived from
single-component fits to the observed surface-brightness distribution
and therefore trace light-weighted, not mass-weighted, galaxy
structure.
In the presence of radial mass-to-light ratio gradients, the stellar
mass distribution can be more centrally concentrated than the observed
light distribution, particularly in star-forming galaxies where the
outer disc contributes disproportionately to the total luminosity
(e.g. \citealt{Mendel2014}).
As a result, light-weighted Sérsic fits may bias some galaxies towards
more disc-like classifications than would be inferred from their
underlying stellar-mass profiles.
This effect is expected to be most relevant for actively star-forming
systems and should be kept in mind when interpreting morphology
fractions and morphology-dependent trends in the present work.

Galaxies were classified into three broad morphological classes:
disc-dominated systems (low Sérsic index, $n < 2$), intermediate
systems ($2 \leq n < 4$), and bulge-dominated systems (high Sérsic
index, $n \geq 4$).
These thresholds are intended to provide a simple separation between
low-concentration and high-concentration systems, but we note that the
high-Sérsic population is itself heterogeneous and may include
classical bulges, ellipticals, and lenticular systems with a wide range
of structural concentrations.
Our primary analysis focuses on the disc- and bulge-dominated
populations, while the intermediate class is retained for completeness
but not interpreted in detail. We use the term `bulge-dominated' as a shorthand for the high-concentration
($n \geq 4$) population selected by these single-component fits. This
classification does not imply that these systems are pure spheroids or that
they lack a significant disc component.

Only galaxies with reliable Sérsic measurements were included in
the morphological analysis.
Specifically, we required successful single-component Sérsic
model fits with well-constrained Sérsic indices, excluding
sources flagged as unresolved (PSF-like), objects with failed or
poor-quality fits, and cases with unphysical or highly uncertain
Sérsic values.
This selection ensured that the adopted Sérsic index provided a robust
first-order description of the galaxy light concentration, while not
attempting to capture the full structural complexity of the galaxy or
its underlying stellar-mass distribution.
Table~\ref{tab:morph_counts} summarises the number of galaxies in
each field with available and reliable morphological
information. Based on our classification criteria there are 17\,721 disc-dominated systems, 3\,252 bulge-dominated galaxies and 561 sources classified as intermediate.

The intermediate morphological class likely represents a heterogeneous
population. Such systems may include disc galaxies with prominent bulges,
early-type spirals, galaxies undergoing morphological
transformation, or systems affected by projection effects and
fitting uncertainties. As a result, this class does not correspond to a single, well-defined physical morphology.

\subsection{sSFR binning scheme}
\label{sec_binning}

To explore environmental trends as a function of star formation
activity, we divided galaxies into bins of sSFR.
Rather than adopting fixed sSFR thresholds, we employed a
quantile-based binning scheme constructed from the combined
galaxy sample across both fields.
This approach ensured comparable statistics in each bin and
minimises potential biases arising from field-to-field
differences in the underlying sSFR distributions.

A key motivation for this choice is to enable a controlled
comparison between disc- and bulge-dominated galaxies.
In particular, one of our primary goals was to define sSFR bins
such that galaxies of different morphology occupying the same
sSFR bin have similar $M_\star$ distributions.
This consideration is important because previous galaxy
clustering studies have reported a dependence of DMH mass on $M_\star$, even at fixed star formation activity
(e.g. \citealt{Coil2017, Mountrichas2019}).
By adopting a quantile-based sSFR scheme, we ensured that the
median $M_\star$ of disc- and bulge-dominated galaxies
within the same sSFR bin differ by less than $\sim$0.3 dex,
thereby reducing the likelihood that residual $M_\star$
differences drive the observed clustering trends.

The galaxy population was split into five sSFR bins spanning the full
range of star formation activity sampled by the data, from the
lowest-sSFR systems (bin 1) to the most actively star-forming systems
(bin 5).
By construction, each bin contained an approximately equal number of
galaxies in the combined sample. As a consequence of this quantile-based definition, the clustering
uncertainties are expected to be of similar order across bins, since
each subsample contains comparable numbers of galaxies.
At the same time, the highest-sSFR bin is open-ended and therefore spans
a broader range of star formation activity than the intermediate bins,
so its median value should be interpreted as representative of a wider
underlying distribution.
This allowed us to probe environmental trends continuously across the
dynamic range of star formation activity while maintaining sufficient
statistics for clustering measurements in each bin.

\begin{figure*}
\centering
\includegraphics[height=6.cm]{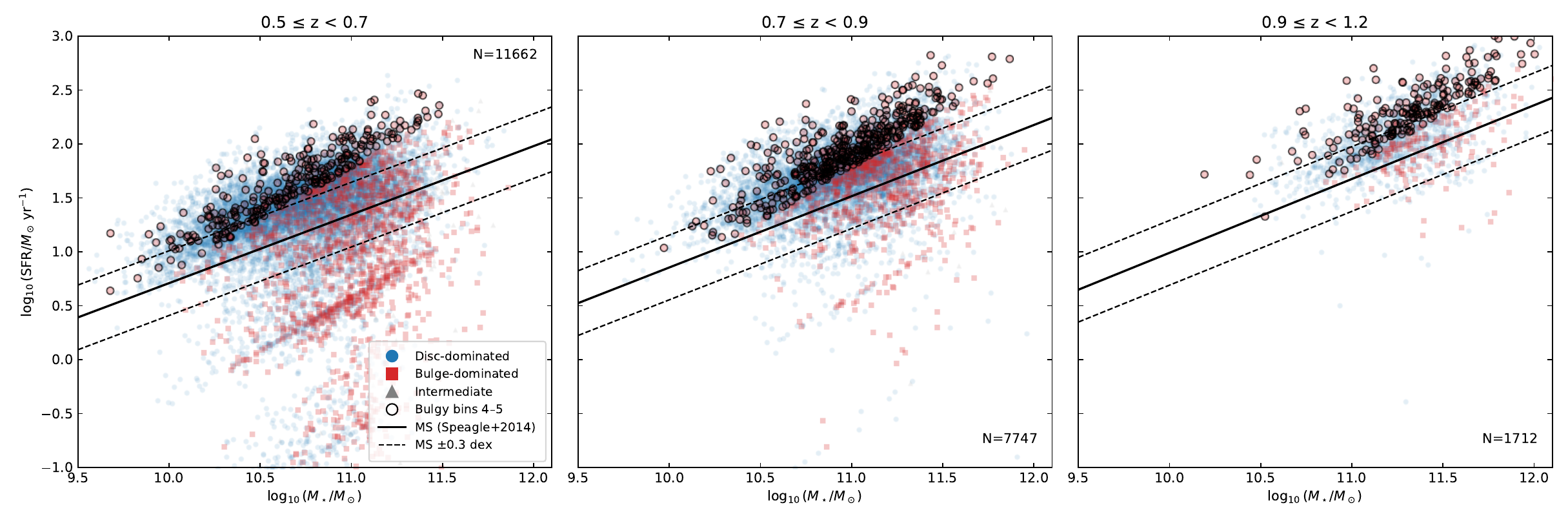}
\caption{$M_\star$--SFR plane for the full galaxy sample in three
redshift slices.
Blue, red, and grey points show disc-dominated, bulge-dominated, and
intermediate-morphology galaxies, respectively.
The solid black line indicates the evolving star-forming main sequence
from \citet{Speagle2014}, while the dashed lines mark
$\pm 0.3$ dex around it.
Bulge-dominated galaxies belonging to the two highest-sSFR bins are
highlighted with black open circles.
The figure shows that these systems lie mostly on or modestly above the
evolving main sequence, rather than representing extreme outliers.}
\label{fig:mstar_sfr_ms}
\end{figure*}

To account explicitly for the redshift evolution of the star-forming
main sequence across the interval $0.5 \le z \le 1.2$, we computed
$\Delta{\rm MS} = \log {\rm SFR} - \log {\rm SFR}_{\rm MS}$ for all
galaxies, adopting the evolving main-sequence relation of
\citet{Speagle2014}.
Figure~\ref{fig:mstar_sfr_ms} shows the $M_\star$--SFR plane in three
redshift slices, with the corresponding main-sequence relation
overplotted.
The figure illustrates that the galaxies occupy the expected locus of
star-forming systems at each redshift, while the bulge-dominated
galaxies in the two highest-sSFR bins lie mostly on or moderately above
the evolving main sequence rather than representing extreme outliers.

We further verified that the current quantile-based sSFR bins map
monotonically onto $\Delta{\rm MS}$ for both disc- and bulge-dominated
galaxies.
In particular, bins~2--5 form a well-ordered sequence in relative
star formation activity, while the lowest-sSFR bin is broader and more
heterogeneous, containing a mixture of systems below or near the main
sequence.
For this reason, we retained the original sSFR-based binning for the
clustering analysis, since it was designed to minimise $M_\star$
differences while preserving sufficient statistics, but in the results
section we present the halo-mass trends as a function of the median
$\Delta{\rm MS}$ of each bin rather than raw sSFR.

\begin{figure}
\centering
\includegraphics[width=\columnwidth]{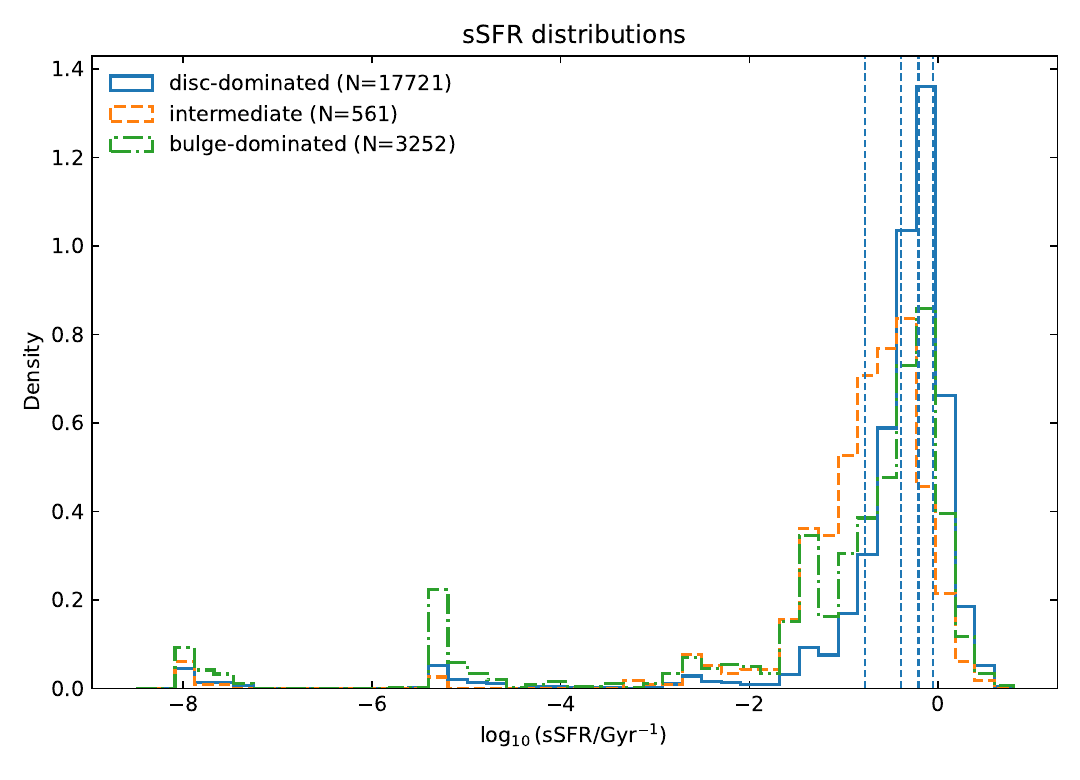}
\caption{sSFR distributions for disc-, intermediate-, and bulge-dominated galaxies. The curves show normalised density distributions, and the vertical dashed lines indicate the quantile-based sSFR bin boundaries adopted for the clustering analysis.}
\label{fig:ssfr_dist}
\end{figure}

Figure~\ref{fig:ssfr_dist} illustrates the sSFR distributions for
disc-, intermediate-, and bulge-dominated galaxies, together with the
adopted quantile boundaries.
Disc- and bulge-dominated systems exhibit partially overlapping sSFR
distributions, with differences that are less pronounced than might be
expected from a purely morphology-driven picture.
This broad behaviour is nevertheless consistent with the expected global
separation between more actively star-forming, lower-concentration
systems and more weakly star-forming, higher-concentration galaxies,
while also highlighting the substantial overlap between the two
populations.
This reflects the fact that morphology and star formation activity are
correlated but not uniquely linked, and that galaxies with different
structural properties can occupy similar star formation regimes,
particularly when morphology is characterised using single-component
structural indicators.

We note that the present sample does not exhibit a strong bimodality in
sSFR, and our quantile-based bins are therefore not intended to define a
strict division between quenched and star-forming galaxies.
Instead, the quantile-based bins provide a relative ordering in
star formation activity that is optimised for a controlled clustering
comparison across morphological classes, and which can also be
interpreted consistently in terms of $\Delta{\rm MS}$.

Galaxies classified as having intermediate Sérsic indices ($2 \leq n < 4$) show sSFR distributions that are not intermediate between those of disc- and bulge-dominated systems.
Instead, they tend to populate lower sSFR regimes and exhibit a broad range of properties.
This behaviour likely reflects the heterogeneous nature of this class, which may include early-type spirals, systems undergoing morphological transformation, galaxies in the
process of quenching, and systems affected by uncertainties in Sérsic
fitting.

In addition, the number of intermediate galaxies is significantly smaller than that of the other morphological classes, particularly when subdivided into sSFR bins.
The combination of low statistics and intrinsic diversity limits the physical interpretability of their clustering measurements.
For these reasons, we did not include intermediate-morphology
galaxies in the clustering analysis and focused our interpretation on the more robust disc- and bulge-dominated samples.

Table~\ref{tab:ssfr_bins} reports the number of galaxies in each
sSFR bin for the different morphological classes.
The resulting binning scheme provides a robust framework for
isolating the role of halo mass in regulating star formation
across galaxy morphologies.

\begin{table}
\centering
\caption{Number of galaxies in each field and the subset with available (reliable) Sérsic index measurements from the DESI morphology catalogues.}
\label{tab:morph_counts}
\begin{tabular}{lcc}
\hline
Field & $N_{\rm gal}$ & $N_{\rm gal}$(with Sérsic) \\
\hline
VIPERS      & 2428  & 2411 \\
Stripe 82 & 19141 & 19123 \\
\hline
Total & 21569 & 21534 \\
\hline
\end{tabular}
\end{table}

\begin{table*}
\centering
\caption{Galaxy properties and inferred halo masses in bins of
specific star formation rate for disc- and bulge-dominated
galaxies.}
\label{tab:ssfr_bins}
\begin{tabular}{lcccccccc}
\hline
 & \multicolumn{4}{c}{Disc-dominated (Sérsic $n<2$)} &
   \multicolumn{4}{c}{Bulge-dominated (Sérsic $n\ge4$)} \\
\cline{2-5}\cline{6-9}
sSFR bin range
& $N$ & $\log M_\star$ & $\log \mathrm{SFR}$ & $\log M_{\rm h}$
& $N$ & $\log M_\star$ & $\log \mathrm{SFR}$ & $\log M_{\rm h}$ \\
\hline
$[-8.09,\,-0.79)$ 
& 2769 & $11.03 \pm 0.26$ & $0.38 \pm 1.51$ & $12.87 \pm 0.28$
& 1302 & $11.20 \pm 0.24$ & $0.40 \pm 1.59$ & $13.26 \pm 0.40$ \\

$[-0.79,\,-0.40)$ 
& 3554 & $11.03 \pm 0.20$ & $1.47 \pm 0.21$ & $12.98 \pm 0.13$
&  596 & $11.15 \pm 0.20$ & $1.54 \pm 0.21$ & $13.53 \pm 0.31$ \\

$[-0.40,\,-0.21)$ 
& 3734 & $10.90 \pm 0.20$ & $1.60 \pm 0.21$ & $12.98 \pm 0.13$
&  484 & $11.05 \pm 0.22$ & $1.76 \pm 0.21$ & $12.91 \pm 0.28$ \\

$[-0.21,\,-0.06)$ 
& 3826 & $10.77 \pm 0.21$ & $1.64 \pm 0.22$ & $13.17 \pm 0.17$
&  441 & $11.03 \pm 0.22$ & $1.89 \pm 0.22$ & $12.15 \pm 0.24$ \\

$[-0.06,\,0.71)$  
& 3838 & $10.63 \pm 0.22$ & $1.73 \pm 0.23$ & $13.05 \pm 0.16$
&  429 & $10.90 \pm 0.29$ & $2.02 \pm 0.27$ & $12.28 \pm 0.27$ \\
\hline
\end{tabular}
\tablefoot{The sSFR bins are defined using quantiles of the combined galaxy
sample with reliable Sérsic measurements. 
Stellar masses and SFRs are reported as medians of
$\log_{10}(M_\star/M_\odot)$ and
$\log_{10}(\mathrm{SFR}/(M_\odot\,\mathrm{yr}^{-1}))$,
with uncertainties corresponding to half the interquartile range
$(Q_{75}-Q_{25})/2$.
Halo masses $\log M_{\rm h}$ are in units of
$\log_{10}(h^{-1}M_\odot)$ and are derived from clustering measurements;
quoted uncertainties correspond to the $1\sigma$ errors propagated from
the bias–halo mass relation.
}
\end{table*}

\section{Clustering analysis and results}
\label{sec_results}

In this section, we present the clustering measurements and the
corresponding inferred DMH masses for the full galaxy
sample, as well as for subsamples split by morphology as a function of
$\Delta{\rm MS}$, as defined in Section~\ref{sec_analysis}. The clustering methodology follows the approach adopted in Papers~I and
II, to which we refer for full technical details.

\subsection{Clustering analysis}

We measured the large-scale clustering of galaxies to infer the typical masses of their host DMHs. The clustering analysis followed the same methodology adopted in Papers~I and II, as well as in \citet{Mountrichas2016} and \citet{Mountrichas2019}, and we refer the reader to those works for a detailed description of the estimators, uncertainty estimation, and halo mass inference.
The clustering measurements were performed using the redshift information
available for the two survey fields, namely spectroscopic redshifts in
VIPERS and primarily spectroscopic, supplemented by high-quality
photometric redshifts where required, in Stripe~82, following the same
approach as in Papers~I and II.

Briefly, we computed the projected two-point cross-correlation
function, $w_p(r_p)$, between each galaxy subsample and a reference galaxy
sample drawn from the same field \citep[e.g.][]{Mountrichas2009b}. The use of cross-correlations, rather than auto-correlations, allows robust clustering measurements to be taken even for subsamples with
limited statistics \citep[e.g.][]{Mountrichas2009a, Hickox2009, Donoso2010,  Krumpe2010a, Miyaji2011, Mountrichas2012, Mountrichas2013, Shen2013, Georgakakis2014, Krumpe2012, Mendez2016, Shirasaki2016, Krumpe2018, Georgakakis2019}. The correlation functions were measured on large scales ($\geq 4\,Mpc$), where the clustering signal is dominated by the two-halo term and is
therefore directly related to the linear bias of the galaxy
population.

Uncertainties on the correlation functions were estimated using
the jackknife resampling technique \citep{Ross2008}, following the same field
subdivision scheme as in Papers~I and II. $w_p(r_p)$ was fitted on
large scales to derive the linear bias of each galaxy subsample.
The measured bias values were then converted into characteristic
DMH masses using standard bias--halo mass relations
derived from numerical simulations (e.g. \citealt{Sheth2001,
Tinker2010}), providing an estimate of the typical halo mass
hosting each population.

The clustering measurements were performed independently in the
VIPERS and Stripe~82 fields. Halo masses derived in the two fields were found to be consistent within the uncertainties and were combined to obtain the final results presented in this section. The resulting halo mass estimates are presented as a function of $\Delta{\rm MS}$ for the full galaxy population and, separately, for disc- and bulge-dominated systems.

\subsection{Halo mass versus $\Delta{\rm MS}$ split for the full galaxy population}
\label{sec_res_full_sample}

\begin{figure}
\centering
\includegraphics[width=0.95\columnwidth, height=7.cm]{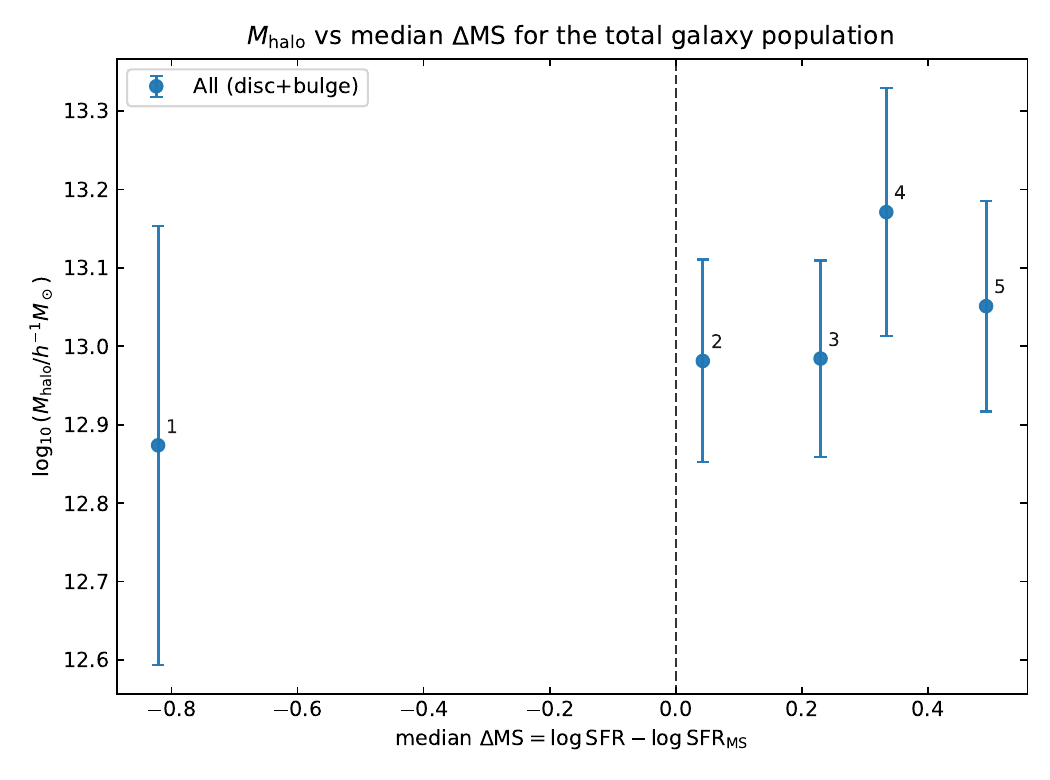}
\caption{Dark-matter halo mass as a function of the median
$\Delta{\rm MS}$ for the total galaxy population.
The measurements include all non-active galaxies with reliable Sérsic
index estimates, combining disc- and bulge-dominated systems.
Each point corresponds to one of the original quantile-based sSFR bins,
labelled from 1 to 5.
The x axis gives the median
$\Delta{\rm MS} = \log {\rm SFR} - \log {\rm SFR}_{\rm MS}$ of the
galaxies in that bin, while the y axis shows the characteristic halo
mass inferred from clustering.
Vertical error bars indicate the $1\sigma$ uncertainties propagated from
the bias--halo mass relation.}
\label{fig:dmhm_dms_full}
\end{figure}

We first examine how the typical DMH mass varies with the median
$\Delta{\rm MS}$ of the full galaxy population, prior to any splitting
by morphology. This analysis includes all non-AGN galaxies with reliable Sérsic index measurements, as described in Section~3 and summarised in
Table~\ref{tab:morph_counts}. Restricting the sample to galaxies with available morphological information ensures homogeneity with the analysis presented in the
subsequent subsections.

Figure~\ref{fig:dmhm_dms_full} shows the inferred DMH mass as a function
of the median $\Delta{\rm MS}$ for the combined galaxy sample.
The points correspond to the original quantile-based sSFR bins, but are
now positioned according to the median offset of each bin from the
evolving main sequence.
Within the range of $\Delta{\rm MS}$ probed by our data, the
characteristic halo mass remains approximately constant, with no strong
systematic trend across the full galaxy population.
The typical halo masses are of order
$\log(M_{\rm h}/h^{-1}M_\odot) \sim 12.8$--$13.1$, consistent across the
five bins. We note that the current bins form a monotonic sequence in
$\Delta{\rm MS}$, although the lowest-sSFR bin is broader and more
heterogeneous than the others, as discussed in Section~3.2.

At first glance, this result may appear to differ from previous studies
that reported a negative relation between halo mass and sSFR for galaxy
samples that included a broader mixture of inactive and AGN-hosting
systems (e.g. \citealt{Coil2017, Mountrichas2019}).
However, this apparent difference is naturally explained by sample
definition.

A key distinction between the VIPERS galaxy sample used in
\citet{Mountrichas2019} and the galaxy samples adopted in this series of
papers is that here we explicitly aim to construct a clean
non-active galaxy sample.
For this reason, we excluded not only X-ray selected AGNs but also systems
identified as AGN candidates through SED decomposition.
This criterion preferentially removes a fraction of bulge-dominated
galaxies from the parent sample, and therefore changes the morphological
mix of the resulting galaxy population.
Specifically, galaxies with an AGN fraction greater than 0.2 were removed
from the parent galaxy samples. Applying this criterion removed approximately 45\% of the original
VIPERS galaxy sample (see Section~3.2 of Paper~I).

Cross-matching the \citet{Mountrichas2019} galaxy sample with the DESI
morphological catalogues shows that the majority of galaxies excluded by
the AGN fraction cut are classified as bulge-dominated systems based on
their Sérsic indices. This behaviour is expected, since bulge-dominated galaxies host, on average, more massive black holes and are therefore more likely to be identified as active through SED-based diagnostics, while AGNs hosted by disc-dominated galaxies are more easily diluted by ongoing star formation.

As a consequence, the galaxy samples used in the present work have a
substantially lower fraction of bulge-dominated systems compared to the
sample analysed in \citet{Mountrichas2019}.
While bulge-dominated galaxies accounted for approximately
40--45\% of the total galaxy population in that study, they represent
only about 15\% of the galaxies in the current sample.
As we show in the following subsection, this morphology-dependent effect
of the AGN exclusion plays a key role in shaping the halo-mass trends when galaxies are considered as a single population.

\subsection{Halo mass versus $\Delta{\rm MS}$ split by morphology}

\begin{figure}
\centering
\includegraphics[width=\columnwidth]{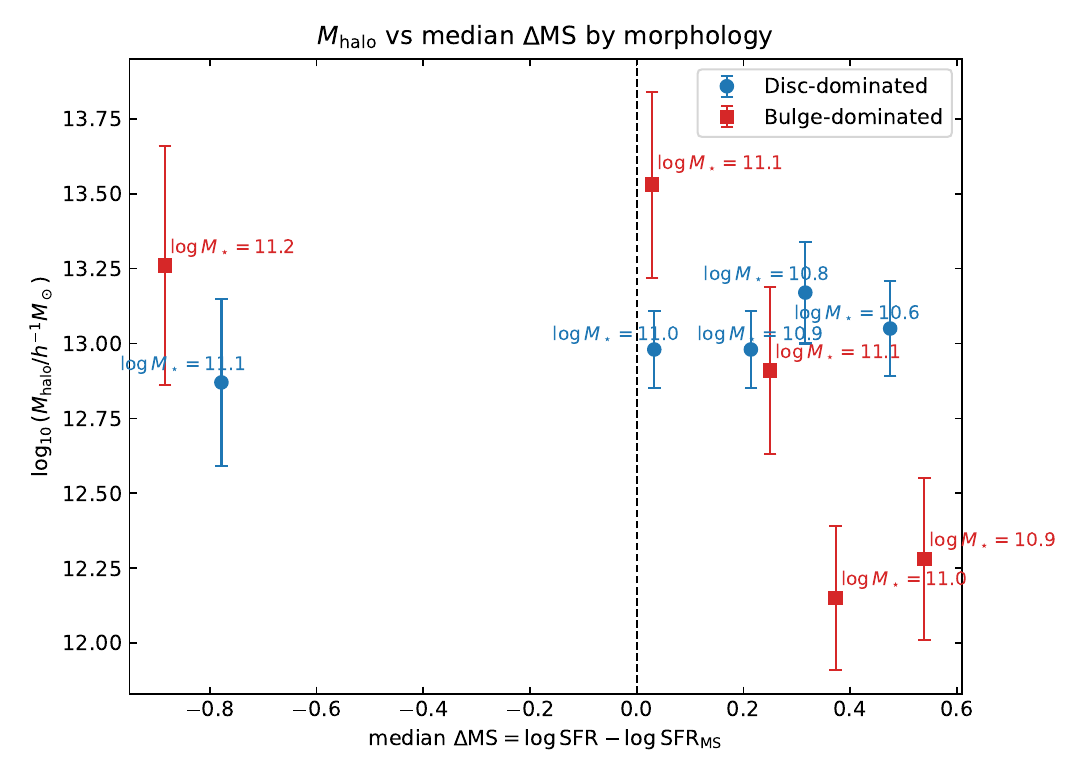}
\caption{DMH mass as a function of median
$\Delta{\rm MS}$ for disc-dominated and bulge-dominated galaxies.
Disc-dominated galaxies are shown in blue and bulge-dominated galaxies
in red.
Each point corresponds to one of the original quantile-based sSFR bins,
but is plotted at the median
$\Delta{\rm MS} = \log {\rm SFR} - \log {\rm SFR}_{\rm MS}$ of the
galaxies in that bin.
The y axis shows the characteristic halo mass inferred from clustering,
and vertical error bars indicate the $1\sigma$ uncertainties propagated
from the bias--halo mass relation.
The text labels report the median stellar mass,
$\log(M_\star/M_\odot)$, of the galaxies in each subsample.
The vertical dashed line marks the evolving star-forming main sequence
($\Delta{\rm MS}=0$).}
\label{fig:dmhm_dms_morph}
\end{figure}

Figure~\ref{fig:dmhm_dms_morph} presents the inferred DMH mass as a
function of the median $\Delta{\rm MS}$ in each bin, separately for
disc-dominated and bulge-dominated galaxies.
The halo masses are derived from clustering measurements in each
original sSFR bin, while the x axis values correspond to the median
offset of the galaxies in that bin from the evolving star-forming main
sequence.

It is important to emphasise that the bulge-dominated points at
$\Delta{\rm MS}>0$ do not represent the median star formation activity of
the bulge-dominated population as a whole. Because the subsamples were
defined using quantiles of the sSFR distribution, these points correspond
specifically to the high-sSFR tail of the $n \geq 4$ population. Their
location above the main sequence therefore does not imply that
bulge-dominated galaxies in general preferentially occupy this regime.

A clear difference emerges between the two morphological classes.
Disc-dominated galaxies exhibit an approximately flat relation between
halo mass and $\Delta{\rm MS}$ across the full range probed in this
analysis.
Within the uncertainties, disc-dominated systems reside in halos of
similar mass irrespective of their relative position with respect to the
main sequence.
In contrast, bulge-dominated galaxies at different $\Delta{\rm MS}$
inhabit halos of different typical mass.
The measurements suggest a transition from lower halo masses at the
high-$\Delta{\rm MS}$ end to higher, mutually consistent halo masses
towards lower $\Delta{\rm MS}$.

Several caveats are important for the interpretation of these trends.
First, the dynamic range in SFR probed by disc-dominated galaxies is
relatively narrow compared to that of bulge-dominated systems.
As shown in Table~\ref{tab:ssfr_bins}, even in the highest-sSFR bins,
the median SFR of disc-dominated galaxies reaches values of order
$\sim$50~$M_\odot\,\mathrm{yr}^{-1}$, whereas bulge-dominated systems
extend to higher SFRs, with median values approaching
$\sim$100~$M_\odot\,\mathrm{yr}^{-1}$.
As a consequence, the approximately flat halo-mass--$\Delta{\rm MS}$
relation observed for disc-dominated galaxies may partly reflect the
limited range of star formation activity sampled by this population.
Testing whether this weak halo-mass variation persists in
disc-dominated galaxies at the highest star formation rates will require
samples that probe a broader dynamic range in star formation activity
than is available here.
Second, the apparent flattening at the high-$\Delta{\rm MS}$ end for
bulge-dominated galaxies should be interpreted with caution, given the
relatively small number of bulge-dominated galaxies in these bins and
the broad, open-ended nature of the upper sSFR tail.

Third, $M_\star$ is known to influence galaxy clustering and halo mass,
with several studies reporting a dependence of halo mass on $M_\star$ at
fixed star formation activity (e.g. \citealt{Coil2017,
Mountrichas2019}).
For this reason, the sSFR binning scheme was explicitly designed to
minimise $M_\star$ differences between morphological classes.
As shown in Table~\ref{tab:ssfr_bins}, the median $M_\star$ of
disc- and bulge-dominated galaxies within the same sSFR bin differs by
at most $\sim$0.3 dex.

Importantly, within each morphological class the median $M_\star$
varies only weakly across the bins.
In particular, for bulge-dominated galaxies the increase in halo mass
towards lower $\Delta{\rm MS}$ is accompanied by only a modest increase
in $M_\star$ of 0.3 dex.
This suggests that the observed halo-mass trend is unlikely to be driven
primarily by residual $M_\star$ variations, although a modest
contribution cannot be fully excluded.
Instead, it reflects a genuine change in the typical halo environments
of bulge-dominated galaxies as a function of star formation activity.

The morphology-dependent trends discussed above also provide a natural
explanation for the approximately flat halo mass--$\Delta{\rm MS}$ relation observed
for the total galaxy population (section~\ref{sec_res_full_sample}).
The exclusion of systems with significant AGN contribution, combined
with the preferential removal of bulge-dominated galaxies from the
parent samples, leads to a galaxy population dominated by
disc-dominated systems.
As a result, the total galaxy sample is weighted towards a population
that exhibits little variation in halo mass with $\Delta{\rm MS}$, thereby masking
the increasing halo-mass trend seen for bulge-dominated galaxies when
all morphologies are considered together.

Taken together, these results demonstrate that the relation between halo
mass and star formation activity is strongly shaped by galaxy
morphology.
While disc-dominated galaxies show little variation in halo mass across
$\Delta{\rm MS}$, bulge-dominated systems exhibit a clear dependence,
with galaxies closer to or below the main sequence residing in more
massive halos than their counterparts at the high-$\Delta{\rm MS}$ end.
A physical interpretation of these trends is discussed in the following
section.

\section{Discussion}
\label{sec_discussion}

In this work we investigate the connection between star formation activity and DMH mass as a function of galaxy morphology, focusing exclusively on non-active galaxies.
By combining DESI-based morphological classifications with clustering measurements across quantile-based sSFR bins, which we interpret in terms of the median $\Delta{\rm MS}$ of each bin, we are able to disentangle, at least to first order, the relative roles of halo-driven and morphology-driven quenching mechanisms.

\subsection{Halo mass versus $\Delta{\rm MS}$ at fixed morphology}

Our main result is not that one morphological class becomes cleanly
quenched once it crosses a nominal halo-mass threshold. Rather, the
striking aspect of Figure~4 is that all subsamples already inhabit
massive halos, typically of order
$\log(M_{\rm h}/h^{-1}M_\odot)\sim 13$, while many of the galaxies
remain on or above the star-forming main sequence. Halo mass therefore
does not map cleanly onto quiescence in the present sample. At the same
time, the relation between halo mass and $\Delta{\rm MS}$ differs
markedly between bulge-dominated and disc-dominated galaxies, indicating
that the coupling between halo environment and star formation activity
depends strongly on morphology.

For bulge-dominated systems, we find evidence that galaxies at
different $\Delta{\rm MS}$ inhabit halos of different typical mass.
In particular, the measurements suggest a transition from lower halo
masses at the high-$\Delta{\rm MS}$ end to higher, mutually consistent
halo masses towards lower $\Delta{\rm MS}$. Presenting the results as a
function of $\Delta{\rm MS}$ makes clear that the morphology-dependent
trends are not driven by the global redshift evolution of the
star-forming main sequence, but persist when galaxies are compared
relative to the evolving locus of normal star-forming systems.
However, the fact that even the high-$\Delta{\rm MS}$ bulge-dominated
subsamples already reside in massive halos implies that the results do
not support a simple threshold picture in which crossing a critical halo
mass is by itself sufficient to quench galaxies.

This point is important when comparing our results to recent claims
that halo mass may be the dominant prerequisite for quenching in
central galaxies \citep{ZhaoPeng2025}. Their analysis is restricted to
centrals and uses a different route to halo-mass estimation, whereas our
clustering analysis yields characteristic large-scale halo masses for
broad galaxy subsamples whose intrinsic halo distributions may
themselves be broad. The two results are therefore not directly
one-to-one comparable. Our data suggest that, at least for the massive
galaxies probed here, typical halo mass alone does not determine whether
a galaxy is quenched, even if halo environment still contributes to
setting the conditions under which quenching becomes more likely.

At the same time, the present analysis does not uniquely demonstrate
halo mass as the primary causal driver of quenching, since halo mass is
itself correlated with other central galaxy properties, including bulge
growth and black-hole mass (e.g. \citealt{Woo2013, Bluck2014,
Bluck2016, Piotrowska2022, Goubert2025}). In this context, we note that
Paper~I of this series found no statistically significant large-scale
halo-mass trend with black-hole mass once host-galaxy properties were
controlled, suggesting that a strong one-to-one coupling between halo
mass and black-hole mass is unlikely to be the primary driver of the
morphology-dependent trends reported here.

For bulge-dominated galaxies, the halo masses inferred for the two
lowest-$\Delta{\rm MS}$ bins, which correspond to the two lowest-sSFR
bins of the original quantile-based classification, are consistent
within the uncertainties. This behaviour may indicate that once galaxies
reach the lowest-$\Delta{\rm MS}$ regime sampled in our data, the
characteristic halo mass saturates, while variations in $\Delta{\rm MS}$
reflect differences in quenching timescales and gas-regulation
processes rather than continued halo-mass growth. Such an interpretation
is plausible given that star formation suppression can occur on shorter
timescales than halo mass assembly. We also note that the lowest-
$\Delta{\rm MS}$ bin is intrinsically heterogeneous, encompassing
systems with a range of evolutionary histories, which may further dilute
monotonic trends.

A similar behaviour is observed at the high-$\Delta{\rm MS}$ end for
bulge-dominated galaxies, where the inferred halo masses of the two
highest-$\Delta{\rm MS}$ bins, corresponding to the two highest-sSFR bins
of the original quantile-based classification, are also consistent within
the uncertainties. These subsamples should not be overinterpreted: they
contain only a few hundred galaxies each, and the differences with the
lower-$\Delta{\rm MS}$ bulge bins are at most of moderate statistical
significance.

It is important to stress that the existence of these high-$\Delta{\rm MS}$
systems is not in itself in conflict with the general association between
prominent bulges and reduced star formation activity. The points represent
the high-sSFR tail of the $n \geq 4$ population rather than the
bulge-dominated population as a whole. Moreover, a high single-component
S\'ersic index does not imply the absence of a star-forming disc. Studies
based on bulge--disc decompositions have shown that galaxies with
significant bulge components can still lie in the star-forming population
when substantial star formation remains present in a disc component
(e.g. Dimauro et al. 2022). Because our morphology is based on
single-S\'ersic fits rather than full bulge--disc decomposition, we cannot
determine directly whether the enhanced SFR in these high-$\Delta{\rm MS}$
bulge-dominated systems is primarily associated with a residual or regrown
disc component.

More generally, the non-monotonic behaviour observed for
bulge-dominated galaxies suggests that halo mass does not map uniquely
onto a given level of star formation activity. Instead, halo mass
appears to set a boundary condition that regulates the range of
star formation states accessible to a galaxy, rather than determining
$\Delta{\rm MS}$ in a one-to-one fashion. In this framework, the shift
towards lower $\Delta{\rm MS}$ with increasing halo mass reflects a
stronger coupling between halo environment and star formation state in
bulge-dominated galaxies, while the flattening observed at both the low-
and high-$\Delta{\rm MS}$ ends indicates regimes in which star formation
activity is governed primarily by internal or short-timescale
processes.

In contrast, disc-dominated galaxies exhibit no significant variation in
halo mass across the range of $\Delta{\rm MS}$ probed in this study.
Within the uncertainties, disc-dominated systems reside in halos of
comparable mass regardless of their star formation activity, consistent
with previous observational studies that find a weak dependence of
environment on star formation for disc galaxies (e.g.
\citealt{Skibba2009, Moustakas2013}). This behaviour suggests that,
within the range probed here, star formation activity in discs is less
tightly coupled to halo environment than in bulge-dominated systems.
This may reflect a larger role for internal or secular regulation, or
other short-timescale processes associated with disc structure and gas
supply. While this is qualitatively compatible with the broader idea
that internal structure can regulate star formation, we do not interpret
it as direct evidence for the specific morphological-quenching scenario
of \citet{Martig2009,Martig2013}, in which a dynamically hot central
stellar component stabilises the gas disc against fragmentation and
thereby suppresses star formation.
Taken together, these results suggest a morphology-dependent connection
between halo environment and star formation activity, but they do not
support a simple picture in which either halo mass or morphology alone
is sufficient to determine whether a galaxy is quenched.

\subsection{Relation to halo quenching and morphological
quenching scenarios}

The observed trends do not support a simple picture in which crossing a
critical halo-mass threshold is by itself sufficient to quench galaxies.
Although our subsamples typically inhabit massive halos, many systems
remain on or above the main sequence, indicating that the connection
between halo mass, morphology, and star formation activity is more
complex than implied by simple threshold quenching models.

This does not mean that halo environment is irrelevant. Halo-scale
processes such as virial shock heating, reduced cold-gas accretion, or
maintenance heating of the circumgalactic medium may still help regulate
the long-term availability of fuel for star formation (e.g.
\citealt{Keres2005, DekelBirnboim2006, GaborDave2012}). Our results
suggest, however, that their observational imprint is morphology
dependent. In bulge-dominated galaxies, lower $\Delta{\rm MS}$ is
associated with somewhat larger characteristic halo masses, whereas in
disc-dominated galaxies similarly massive halos coexist with sustained
star formation.

Likewise, the results do not provide straightforward support for a
classical morphological-quenching picture. The presence of a bulge or
high Sérsic index is clearly not by itself sufficient to imply low
star formation activity, since a substantial population of
bulge-dominated galaxies still lies on or above the main sequence.
Conversely, the absence of a strong halo-mass trend in discs does not
show that halo environment is unimportant, only that in this sample it
is not cleanly translated into quiescence.

A more plausible interpretation is that halo mass, morphology, and the
internal regulation of star formation act together rather than through a
single dominant control parameter. In that sense, our results favour a
coupled picture in which halo environment sets broad boundary
conditions, while the detailed star formation state of a galaxy depends
on its internal structure, gas supply, and evolutionary history.
Testing this picture more stringently will require samples that probe a
broader dynamic range in star formation activity, together with better
constraints on bulge--disc structure and on the full halo-mass
distributions of the underlying galaxy populations.

\subsection{Stellar mass considerations}

Stellar mass ($M_\star$) is a known factor influencing galaxy clustering and halo
occupation, and several studies have reported at most a mild dependence
of halo mass on $M_\star$ at fixed galaxy type or star formation
activity (e.g. \citealt{Mountrichas2019}), while others find no
significant dependence once star formation is accounted for (e.g.
\citealt{Allevato2019, Viitanen2019}).
In the present analysis, $M_\star$ effects are unlikely to drive the
observed morphology-dependent trends. As discussed in Section~4.3, the
median $M_\star$ of disc- and bulge-dominated galaxies within the same
bins, originally defined in sSFR and now interpreted through their
median $\Delta{\rm MS}$, differs by less than $\sim$0.3 dex. Moreover,
within each morphological class the median $M_\star$ varies only weakly
across the bins. In particular, for bulge-dominated galaxies the
increase in halo mass towards lower $\Delta{\rm MS}$ is not accompanied
by a corresponding increase in $M_\star$.

At the same time, the absolute scale of the halo masses we recover is
not surprising. The median stellar masses of our subsamples are all of
order $\log(M_\star/M_\odot)\sim 11$, and abundance-matching arguments
would naturally associate such galaxies with host halos of order
$\log(M_{\rm h}/h^{-1}M_\odot)\sim 13$. In this sense, the relatively
stable halo-mass scale inferred from clustering is broadly consistent
with the typical stellar masses of the galaxies in our sample. The key
result is therefore not the absolute halo-mass scale itself, but the
residual morphology-dependent variation around it.

These results suggest that the morphology-dependent trends in halo mass
are unlikely to be driven primarily by residual M$_\star$ effects,
although a modest contribution from weak M$_\star$ variations cannot
be fully excluded. For this reason, we did not apply explicit $M_\star$
matching between morphological classes, as doing so would unnecessarily
reduce the sample size without materially changing the main physical
interpretation. Nevertheless, some coupling between stellar mass, morphology, and star
formation history is inevitable, and future studies with larger samples
and improved morphological information will be able to explore these
dependencies in more detail.

\subsection{Limitations of morphological classification and future
prospects}

A key limitation of the present analysis is the reliance on
single-component Sérsic indices derived from ground-based imaging.
While the Sérsic index provides a useful first-order proxy for galaxy
morphology, it is measured from the observed light distribution and does
not necessarily trace the underlying stellar-mass structure. In the
presence of radial mass-to-light ratio gradients, light-weighted fits
can underestimate the true central mass concentration, particularly in
star-forming galaxies where the outer disc contributes strongly to the
observed luminosity (e.g. \citealt{Mendel2014}). As a consequence, some
galaxies may be classified as more disc-like in light than in mass,
which can affect both the inferred morphology fractions and the
morphology-dependent trends discussed here.

This limitation is likely to be most relevant for actively star-forming
systems and may blur the separation between disc- and bulge-dominated
galaxies, especially at the high-$\Delta{\rm MS}$ end. More generally,
single-component Sérsic fits cannot capture the full structural
complexity of galaxies, such as bars, spiral arms, or multi-component
bulge--disc systems. Projection effects and fitting uncertainties may
further blur the boundaries between morphological classes, particularly
at intermediate redshift. Because homogeneous mass-weighted structural
measurements are not available for the full sample across both fields
and redshifts considered here, we cannot quantify this effect directly
in the present analysis. The reported trends should therefore be
interpreted as referring to light-weighted structural classes.

A further limitation is that the present sample includes both central
and satellite galaxies, and we do not have homogeneous group-based
central and satellite classifications across both fields and the full
redshift range considered here. Likewise, we do not have uniform
measurements of bulge mass, velocity dispersion, or black-hole-mass
proxies for the full sample. As a result, the present analysis cannot
fully disentangle halo mass from other central galaxy properties known
to correlate strongly with quenching. The trends reported here should
therefore be interpreted as morphology-resolved empirical constraints on
the link between halo environment and star formation activity, rather
than as a unique identification of the primary causal quenching
variable. Although we cannot classify centrals and satellites directly,
literature estimates at similar stellar masses suggest that the
satellite fraction is likely modest, especially for the star-forming
population \citep[e.g. see Fig. 13 and Fig. 9 in][respectively]{Tinker2013, Fu2022}, so the present sample is
plausibly dominated by central galaxies.

In addition, this study is restricted to non-active galaxies. Although
extending the analysis to AGN host galaxies would be highly informative,
the fraction of X-ray AGNs with reliable morphological measurements in
current datasets remains limited, as discussed in Paper~II. As a result,
a parallel investigation of halo mass, star formation activity, and
morphology for AGN hosts is deferred to future work.

More broadly, recent comparisons between observations and cosmological
simulations have underscored the difficulty of reproducing the joint
dependence of quiescence, halo mass, environment, and AGN activity
within a single framework (e.g. \citealt{YesufBottrell2026}). This
further highlights the value of empirical studies, such as the present
one, that isolate the role of halo environment and galaxy structure
using direct halo-mass estimates from clustering.

Upcoming surveys will substantially alleviate these limitations. In
particular, \textit{Euclid} will provide higher-resolution imaging and
much improved structural information over wide areas, enabling more
reliable morphology measurements and better separation of bulge- and
disc-dominated systems. Combined with larger spectroscopic samples and
improved group catalogues, such datasets will allow the relative roles
of halo mass, galaxy structure, and black-hole growth to be tested much
more directly.

\subsection{Summary of physical interpretation}

Overall, our results support a scenario in which galaxy quenching is governed by different mechanisms depending on morphology.
Bulge-dominated galaxies appear to move towards lower star formation
activity in conjunction with halo-mass growth, consistent with
halo-driven suppression of star formation through mechanisms such as
virial shock heating or reduced cold gas accretion in massive halos.
In these systems, morphology and environment appear to act in concert, with bulge growth and halo mass assembly tracing a common evolutionary pathway towards quiescence.

Disc-dominated galaxies, on the other hand, show little evidence for a
strong coupling between halo mass and star formation activity. The roughly
constant halo masses inferred across the bins indicate that variations in
their star formation activity are not primarily traced by changes in the
large-scale halo environment. This may point to a greater role for internal
or secular regulation and variations in gas supply, although the present
measurements do not identify the responsible mechanism. In particular, we
do not interpret the flat halo-mass--$\Delta{\rm MS}$ relation of
disc-dominated galaxies as direct evidence for classical morphological
quenching.

This morphology-dependent behaviour highlights the importance of jointly considering galaxy structure and environment when interpreting the quenching of star formation.
In particular, analyses based solely on morphology-averaged galaxy samples may miss key physical trends that only emerge once different structural populations are examined separately.

\section{Summary}
\label{sec_summary}

We have investigated the connection between galaxy morphology,
star formation activity, and DMH mass using non-active
galaxies in the VIPERS and Stripe~82 fields at $0.5 \le z \le 1.2$.
By combining DESI-based Sérsic classifications with homogeneous
$M_\star$ and SFR estimates, we examined halo-mass trends for the total
galaxy population and separately for disc- and bulge-dominated systems.
The galaxy samples were divided into quantile-based sSFR bins, which we
then interpreted through the median $\Delta{\rm MS}$ of each bin in
order to account for the redshift evolution of the star-forming main
sequence.

Our main result is that the relation between halo mass and
star formation activity differs markedly between morphological classes.
While the total galaxy population exhibits an approximately flat
halo-mass--$\Delta{\rm MS}$ relation, this behaviour masks distinct
trends when galaxies are split by morphology.
Disc-dominated galaxies reside in halos of similar mass across the full
range of $\Delta{\rm MS}$ probed, whereas bulge-dominated galaxies
occupy lower-mass halos at the high-$\Delta{\rm MS}$ end and higher,
mutually consistent halo masses towards lower $\Delta{\rm MS}$.
This behaviour is unlikely to be driven primarily by residual
$M_\star$ differences, which remain modest within and across the bins,
although a secondary contribution from weak stellar-mass variations
cannot be fully excluded.
We note, however, that disc-dominated galaxies in our sample probe a
more limited range of SFR than bulge-dominated systems, and extending
this analysis to higher-SFR discs will be important for testing whether
similar halo-related trends emerge at the extreme end of the
star-forming population.

These findings support a morphology-dependent picture of galaxy
quenching, in which the relation between halo environment and
star formation activity is stronger in bulge-dominated systems than in
disc-dominated galaxies.
They are broadly consistent with halo-related quenching scenarios in
high-concentration galaxies, while also indicating that the present data
do not uniquely isolate halo mass from other correlated central galaxy
properties.
Future surveys with improved morphological information and additional
structural and dynamical diagnostics, such as \emph{Euclid} \citep{Euclid2025Overview} will be required to disentangle
these effects more fully.

\begin{acknowledgements}
We thank the referee for a careful reading of the manuscript and for
constructive comments and suggestions that helped improve the clarity
and quality of this work. GM acknowledges funding from grant PID2021-122955OB-C41 funded by MCIN/AEI/10.13039/501100011033 and by “ERDF/EU”. This work was partially supported by the European Union's Horizon 2020 Research and Innovation programme under the Maria Sklodowska-Curie grant agreement (No. 754510).

\end{acknowledgements}

\bibliography{mybib}
\bibliographystyle{aa}

\end{document}